\documentclass[fleqn,usenatbib]{mnras}
\usepackage{newtxtext,newtxmath}
\usepackage[T1]{fontenc}
\usepackage{ae,aecompl}
\usepackage{todonotes}

\usepackage{graphicx}	
\usepackage{amsmath}	
\usepackage{color}
\usepackage{natbib}

\title[Analysing geomorphological structures in images]{Detecting and analysing geomorphological structures in images of comet 67P/Churyumov-Gerasimenko using Fourier transform} 

\author[Ruzicka et al.]{
Birko-Katarina Ruzicka,$^{1,3}$\thanks{E-mail: ruzicka@mps.mpg.de}
Matthias Schr\"oter,$^{2}$\thanks{E-mail: matthias.schroeter@ds.mpg.de}
Andreas Pack,$^{3}$
\newauthor \hspace{2mm}and Hermann Boehnhardt,$^{1}$
\\
$^{1}$Max-Planck-Institut f\"ur Sonnensystemforschung, Justus-von-Liebig-Weg 3, 37077 G\"ottingen, Germany\\
$^{2}$Max Planck Institute for Dynamics and Self-Organization, 37077 G\"ottingen, Germany\\
$^{3}$Georg-August-Universit\"at G\"ottingen, Geowissenschaftliches Zentrum, Abteilung Isotopengeologie, Goldschmidtstra{\ss}e 1,\\ 
\hspace{1.5mm}37073 G\"ottingen, Germany
}

\date{Accepted XXX. Received YYY; in original form ZZZ}
\pubyear{2019}

\begin{document} 

\label{firstpage} 
\pagerange{\pageref{firstpage}--\pageref{lastpage}}
\maketitle

\begin{abstract} 
We present a method for automatised detection and analysis of quasi-periodic lineament structures from images at pixel-precision. The method exploits properties of the images' frequency domain found by using the Fourier transform. We developed this method with the goal of detecting lineament structures in an image of the Hathor cliff of comet 67P/Churyumov-Gerasimenko, which are caused by layerings and furrows in the nucleus material. Using our method, we determined the orientation and wavelength-range of these structures. The detected layering edges have similar orientations, spatial separations of 9-20\,m, and are ubiquitous throughout the image. We suggest that the layerings are a global feature of the comet nucleus that provide information about formation and evolution of comet 67P. The furrows are non-uniformly distributed throughout the image. Their orientation is broadly parallel to the direction of the local gravity vector at the Hathor cliff, with spacings similar to that of the layering structures. The furrows are interpreted as signatures of local down-slope movement of cliff material. We demonstrate that the developed method is broadly applicable to the detection and analysis of various kinds of quasi-periodic structures like geological layering, folding and faulting, and texture analysis in general. In order to facilitate the application of our method, this paper is accompanied by a demo program written in Matlab.
\end{abstract}

\begin{keywords}
techniques: image processing -- methods: data analysis -- comets: general -- comets: individual: 67P/Churyumov-Gerasimenko -- 
\end{keywords}



\section{Introduction}\label{sec:introduction}

Cometary nuclei are among the oldest and least-altered solid bodies in our Solar System. They provide key information about the conditions and physical processes that shaped the Solar System $\sim$4.6 billion years ago. Space missions from the past two decades have revealed great details about the complex shapes, textures, structures, and composition of cometary nuclei, including the observation that the nucleus material might be structured in layerings. These layerings were first suggested for the nuclei of the comets 19P/Borrelly and 81P/Wild 2  \citep[e.g.][]{britt_2004,brownlee_2004}, but data from those missions was limited in its spatial resolution. The first unambiguous morphological evidence of layerings was found on the much clearer images taken by the Deep Impact probe to comet 9P/Tempel 1 \citep{thomas_shape_2007,belton_internal_2007}.

The ESA Rosetta mission delivered more than 70,000 images of the nucleus of comet 67P/Churyumov-Gerasimenko ('comet 67P'). The spatial resolution of the images reaches the sub-metre range, allowing the study of surface features in unprecedented detail. The images revealed smooth plains, flat terraces, and steep cliffs whose geometric configuration indicate a layered internal structure that is global, concentric, and permeating at least several hundred metres into the nucleus  \citep[]{massironi_two_2015,penasa_three_2017,ruzicka_mnras_2018,belton_origin_2018}. The ongoing quest to understand how and when the layerings were formed in the nucleus hinges on a comprehensive understanding of their 3-dimensional structure, which is a subject of great interest in the community. Open questions include the locations of exposure on the comet, the depth of their extent below the surface of the nucleus, their lateral extension, their orientation within the gravity field, and their individual thickness. 

The first step in answering these questions would be the creation of a global map of the layerings. However, conventional manual approaches thus far \citep[e.g.][]{massironi_two_2015,penasa_three_2017,ruzicka_mnras_2018} resulted in maps of limited resolution, as only large-scale morphological features could be identified with confidence. Furthermore, the results were vulnerable to 'confirmation bias', i.e., the observer's expectation based on previous observation. In addition, manual mapping is labour-intensive and does not easily accommodate updates to the data or the process. 

We therefore developed an algorithm that automatically identifies directed, quasi-periodic, linear structures in images. The algorithm analyses images with minimal human bias and thus ensures a high degree of reproducibility (cf. supplementary material). The algorithm produces a 'map' of the linear features within the image, which are further analysed for properties such as orientation and spacing. Possible applications of the algorithm include jointing, cracks, polygons, furrows, and layerings, all of which were found to be ubiquitous on comet 67P \citep[e.g.][]{thomas_2015_morphological,elmaarry_fractures_2015,auger_polygons_2018}. We use this algorithm to present an in-depth analysis of the layerings exposed on the Hathor cliff, one of the regions on the minor lobe of comet 67P. Finally, we briefly assess which of the currently-discussed models of cometary formation would be compatible with our results.

\begin{figure} 
	\includegraphics[width=\columnwidth]{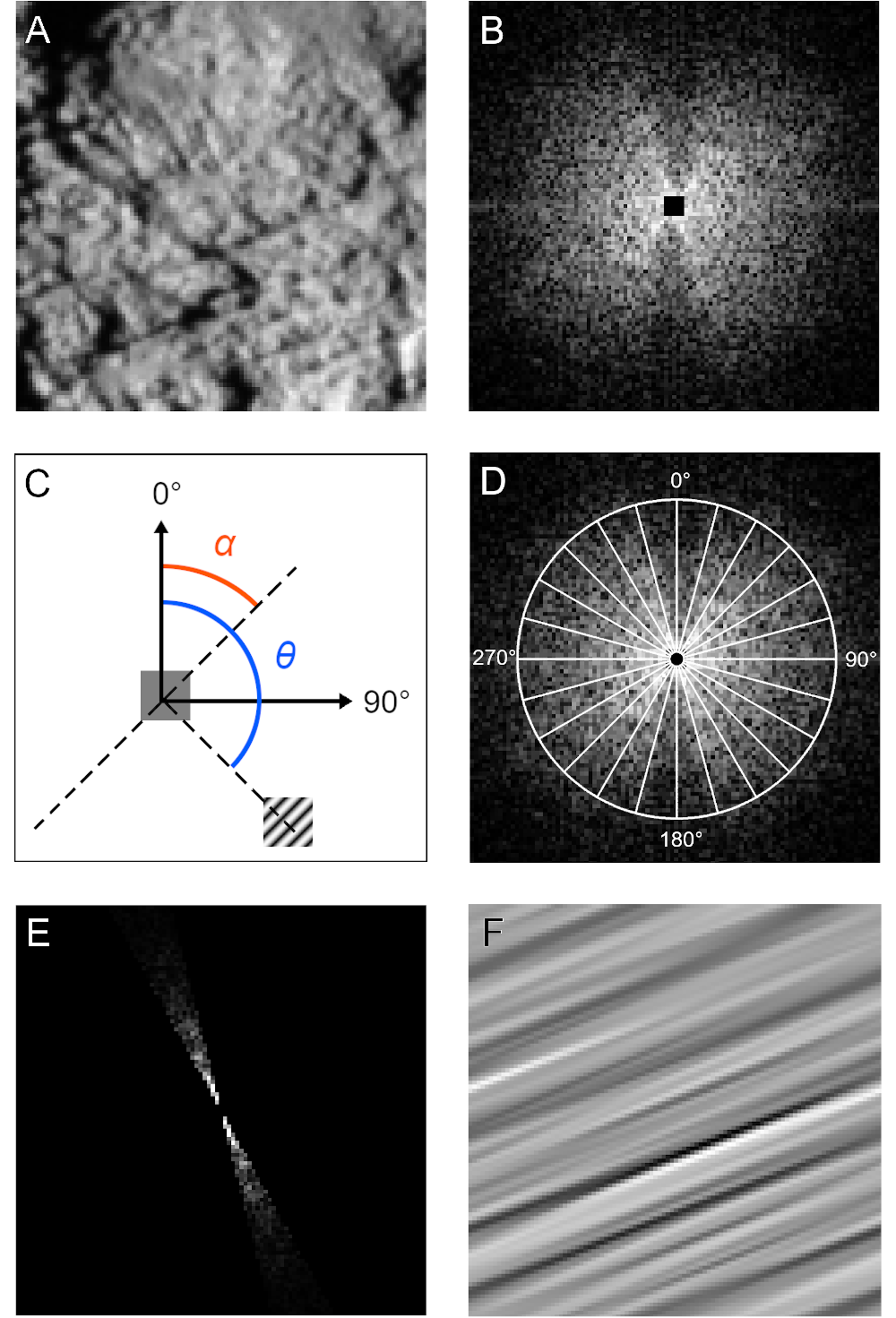}
    \caption{Computational workflow of the Fast Fourier Transform (FFT) approach for analysing layerings on the surface of comet 67P. The workflow is demonstrated on an example image cropped from \autoref{fig:crop_location}. \textbf{A:} the original image; \textbf{B:} the FFT image of the masked image (logarithmic display); \textbf{C:} Directional conventions used in this work, $\alpha$ denotes the orientation of features in the image (panel A), $\theta$ denotes the location of Fourier modes (panel B). Two modes from \autoref{fig:zebra} are shown to give an example; \textbf{D:} schematic drawing of FFT division into sectors (shown with wider sectors and without overlap for clarity); \textbf{E:} reduced FFT, where all sectors are set to zero except one direction of interest (e.g. the lineaments associated with layering edges in panel A), shown with wider sector for clarity; \textbf{F:} backtransformed image using only the FFT information contained within the remaining sector shown in panel E.}
    \label{fig:workflow}
\end{figure}

\begin{figure} 
    \centering
    \includegraphics[width=\linewidth]{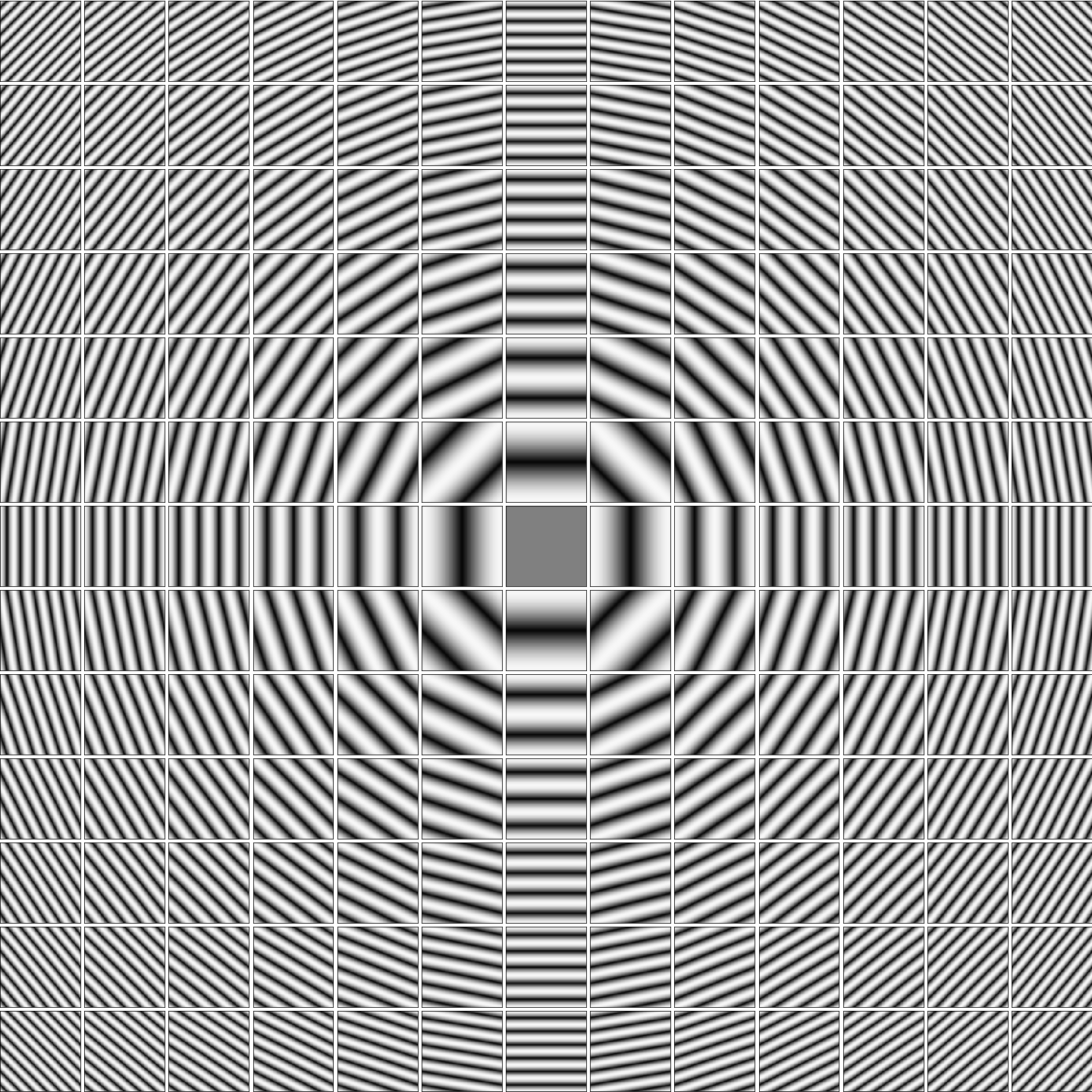}
    \caption{Schematic visualisation of the orientation and wavelength of Fourier modes. Here 13$^2$ = 169 modes are displayed. The number of modes is equal to the number of pixels in the analysed image.}
    \label{fig:zebra}
\end{figure}

\section{Method}\label{sec:method}

\subsection{Fourier Transform for image analysis} \label{sec:fouriertransform}

Fourier transformation is a method to decompose any discretely sampled function into a sum of sine and cosine functions with frequencies which are multiples of a base frequency. The amplitudes of this set of sine and cosine functions are then the Fourier representation of the original signal. Two common ways of using them are a) the analysis of the dominant periodic contribution of the original signal and b) filtering, i.e., the removal of unwanted parts of the signal. The latter works by setting the Fourier amplitudes of the unwanted parts to zero and then using an inverse Fourier transform to create the filtered signal.  

While the the Fourier transformation was first described in 1822 by Jean Baptiste Joseph Fourier, it was the invention of the 'Fast Fourier Transform' (FFT) by Cooley and Tukey in 1965 which dramatically increased the computational efficiency of the algorithm and led to its widespread application in science and engineering. In geology and planetary science, FFT has first been used to analyse one-dimensional elevation data to either find the dominant wavelengths of a structure (for examples see \cite{fowajuh:95,deardorff:19,gehrmann:18}) or to filter the signal in a preprocessing step \citep{hanley:77}.

However, it is also possible to Fourier transform two-dimensional signals such as the height map of an area. The Fourier representation of the signal then consists of sine and cosine waves with different spatial frequencies and orientations; the latter allow to not only determine dominant wavelengths, but also their orientations. 

Such 2D FFTs have been applied to a variety of geology and planetary science problems. Filtering in Fourier space was applied to
microreliefs  \citep{stone:65},
karst structures \citep{harrison:96},
shear zones \citep{pal:06}, 
or ash on ice \citep{nield:13}.  
Dominant wavelengths have been studied by \cite{brook:81} (karst), and \cite{perron:08} (mesa and river landscapes).
The orientation information of the Fourier representation has been used to determine the orientation of fractures \citep{brook:91} and glacial and aeolian lineaments \citep{fowajuh:95}. \cite{davis:17}  studied both wavelength and orientation in karst and coral reef. 

More specialised analysis in Fourier space have been performed
by 
\cite{perron:08} (non-fractal nature of landscapes),
\cite{booth:09} (detection of landslides),
\cite{nield:13}  (roughness analysis),
\cite{bugnicourt:18} (landscape classification), and
\cite{cao:18} (local roughness on the moon).

In this paper we use two-dimensional FFT to identify and verify linear structures in an image of the Hathor cliff. The first step is to compute the FFT of a square image, which has been masked with a window mask. In the resulting frequency domain, the angular distribution of Fourier modes is then analysed for intensity maxima. Computing the image's inverse FFT from the direction of maximum intensity can be used to visually verify whether the signal corresponds to the linear structure of interest in the original image. An overview of this workflow for an example image is shown in \autoref{fig:workflow}.

\subsubsection{Fast Fourier Transform of the selected image} \label{sec:FFT} 

Computing the FFT of an image of size M $\times$ N returns an array called the 'frequency domain' of the image, which consists of M $\times$ N modes. The modes indirectly hold information about dominant length scales and angles in the image (\autoref{fig:zebra}). The 'zero mode' $m_0$, located in the image centre, represents the mean grey value of the image. As it is the most common value, this mode has by far the greatest intensity. With increasing radius \textit{r}, i.e., distance to $m_0$, increasingly smaller wavelengths $\lambda$ are represented. The angle $\theta$ of a mode (which denotes the clockwise-positive angle between the line connecting $m$ and $m_0$, and a vertical line arising upwards from $m_0$) indicates the orientation $\alpha$ in which the linear signal represented by the mode occurs in the image, such that $\alpha = \theta - 90^\circ$ (\autoref{fig:workflow}C).

\begin{figure} 
	\centering
	\includegraphics[width=\linewidth]{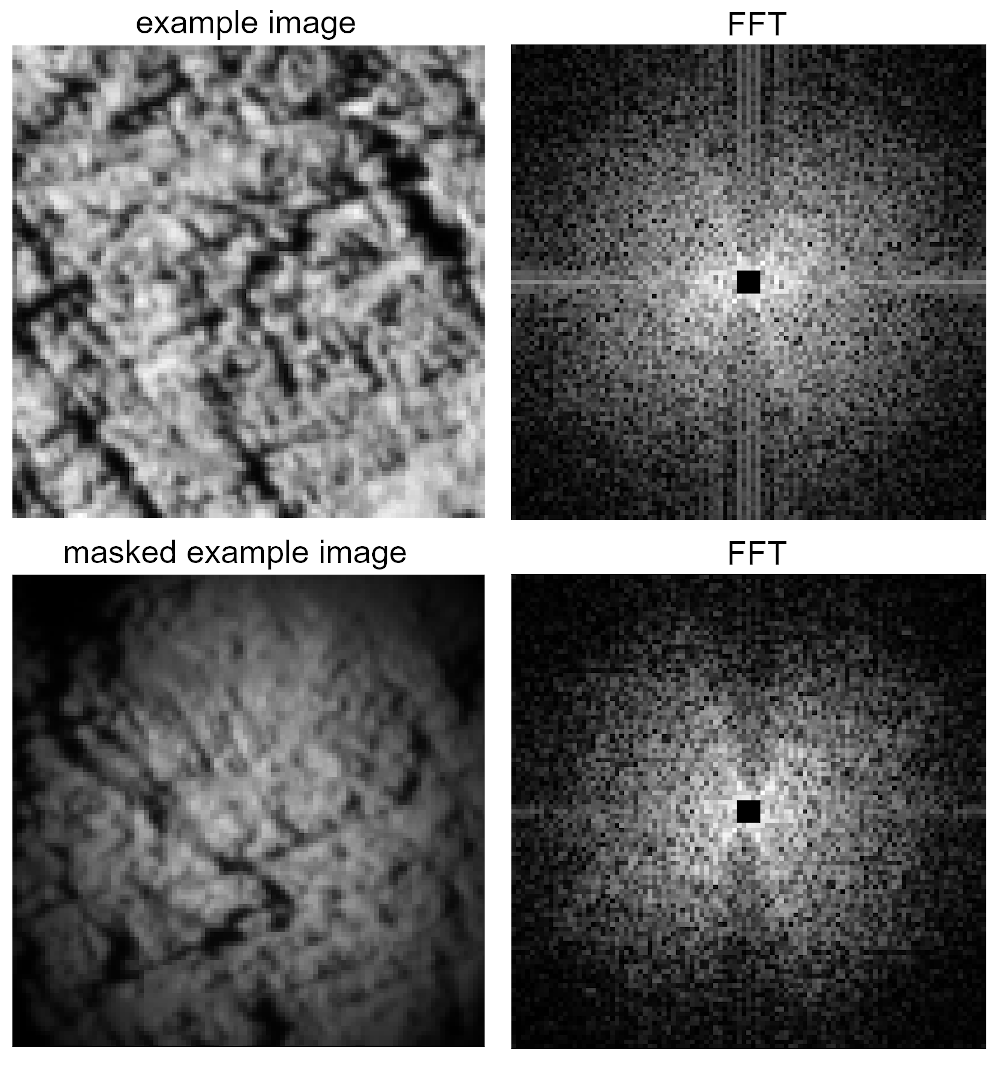}
	\caption[Effect of windowing an image on its FFT]
	{Demonstration of the effect of applying a window mask to an image. \textbf{Top row:} Example image (unmasked) and the corresponding FFT showing leakage lines. \textbf{Bottom row:} Left image: Same image as above, after multiplication with a Gaussian window mask. The image boundaries are now of similar brightness. Note the reduction in area that provides sensitive results during FFT. Right image: No leakage lines in the corresponding FFT.}
	\label{fig:windowing}
\end{figure}

\subsubsection{Masking of the original image} \label{sec:masking} 
 
 When the greyscale-values at opposite image boundaries are notably dissimilar, several parallel lines appear in the FFT image that are crossing its central domain (i.e., around $m_0$) in horizontal and vertical direction (\autoref{fig:windowing}, top row). This effect is called 'leakage' and happens when the image boundaries are incorrectly recognised as edges, as the Fourier transform algorithm is expecting a periodic input signal and therefore repeats the image infinitely. The modes represented by the leakage lines are thus not truly part of the signal, they only appear because short wavelength amplitudes from the image edges have 'leaked' into them. 

The leakage noise can be reduced in the FFT through a process called 'windowing' \citep{bergland:69,harris:78}, i.e., masking the input image with a 2D window function before computing the FFT (\autoref{fig:windowing}, bottom row). We found that a window mask with a Gaussian distribution (1 at the centre, diminishing towards the edges, standard deviation width $\sigma =$ 40\% the length of the image in pixels) sufficiently clears up the frequency domain, while sustaining a high degree of transmission.

\begin{figure} 
	\centering
	\includegraphics[width=0.99\linewidth]{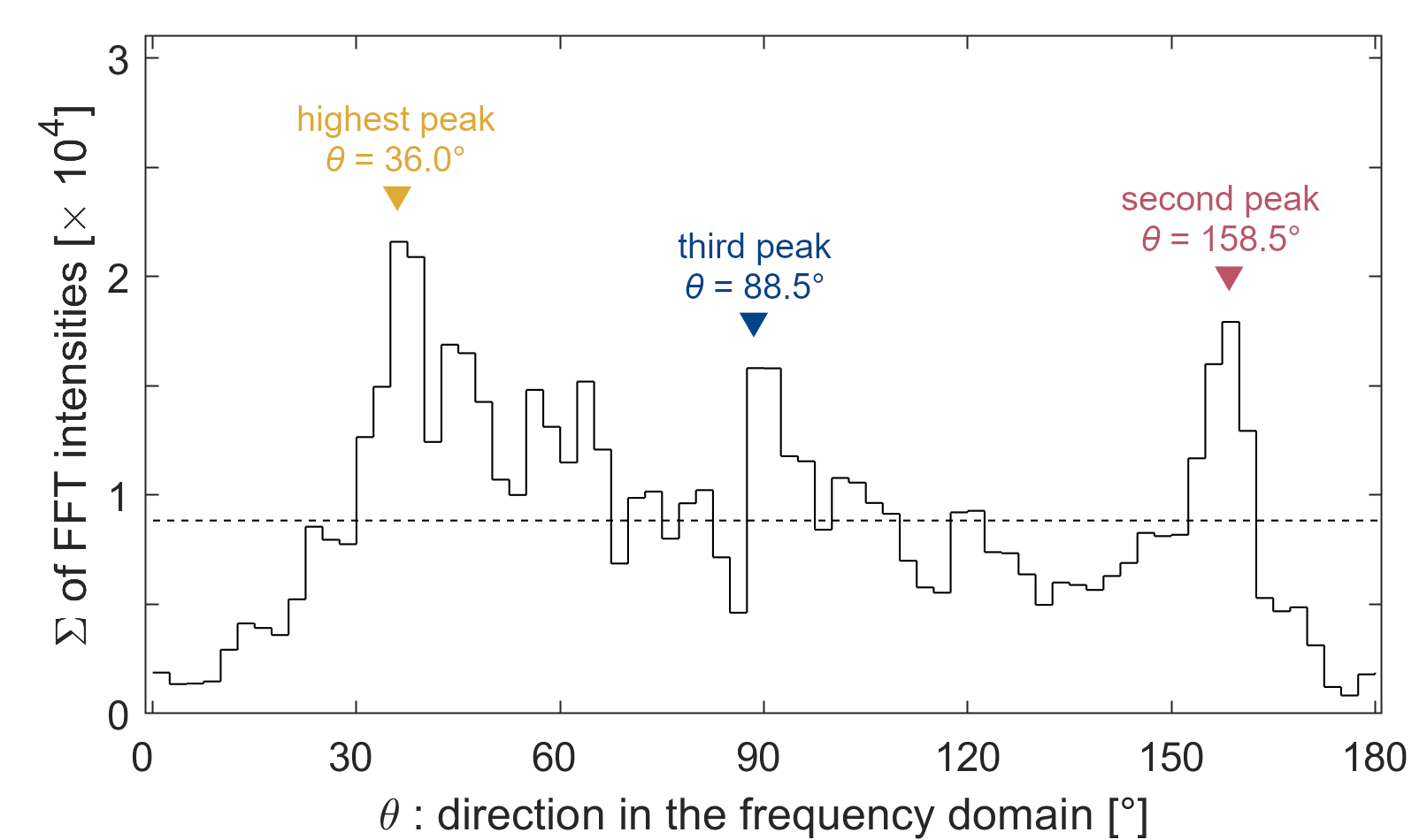}
	\caption{Cumulative FFT-intensities per angular sector ('intensity spectrum') for \autoref{fig:workflow}A; the dashed line indicates the mean intensity of all sectors. Sectors are 5\degr{} wide and overlap by 50\%. Maxima identified by a peak-fitting algorithm are labelled above the histogram. The requirements for peak detection were: Minimum intensity 1.5 $\times$ the mean intensity of all sectors; minimum peak spacing 15\degr{}; minimum peak width 5 sectors (i.e., 12.5\degr{} due to the 50\% overlap). FFT intensity values were squared to enhance the peaks.}.
	\label{fig:intensities_bar}
\end{figure}

\subsubsection{Maxima in the angular distribution of the FFT} \label{sec:peaks} 

The directions $\theta$ of maximum mode intensity represent the directions $\alpha$ of prominent linear structures in the image. We determined the cumulative intensity for each direction $\theta$ in the frequency domain by dividing it into sectors (\autoref{fig:workflow}D) and summing up the intensities of all modes contained in each sector. As each pixel is allocated to only one sector, the resulting intensity sum of the sector ($I_s$) has to be normalised by the number $n^*$ of pixels in each sector to account for variations caused by limited resolution and the square nature of pixels.

The sectors have a minimum ($r_{min}$) and a maximum ($r_{max}$) radius. Parameter $r_{max}$ helps to normalise the number of modes per sector, as otherwise sectors facing a 'corner' of the square image (e.g. at $\theta$ = 45\degr) would include more modes than sectors facing an edge of the domain (e.g. at $\theta$ = 90\degr). As the sector width decreases below the pixel size towards the centre, parameter $r_{min}$ reduces the issue of uniquely allocating modes to one sector.

The cumulative intensity $I_s$ for each sector is found as follows:
\begin{equation}
\forall m(i,j) \text{ with } \theta_{min} \leq \theta(i,j) \leq \theta_{max} \text{ : } I_s = \frac{1}{n^*}
\sum_{r_{min}}^{r_{max}} I(i,j)
\label{eq:sectorsum}
\end{equation}

where $m(i,j)$ is the mode represented by the FFT pixel with coordinates $(i,j)$, and $I(i,j)$ means the intensity of this mode. $I_s$ is a vector with \texttt{nsec} elements, where \texttt{nsec} is the number of sectors. It represents the angular distribution in the frequency spectrum of the studied image. An example spectrum is shown in \autoref{fig:intensities_bar}. Squaring the FFT intensities enhances the amplitudes, which results in a spectrum that shows several groupings of sectors with above-average intensities. A peak-fitting algorithm is used to identify the sectors that have the highest local signal-to-noise ratio in the angular frequency spectrum. To visually verify the presence of repetitive signatures in the landscape of the original image, we created a backtransformed FFT image using only the frequency spectrum of sectors with determined maxima in the angular distribution.

\begin{figure} 
	\centering
	\includegraphics[width=\linewidth]{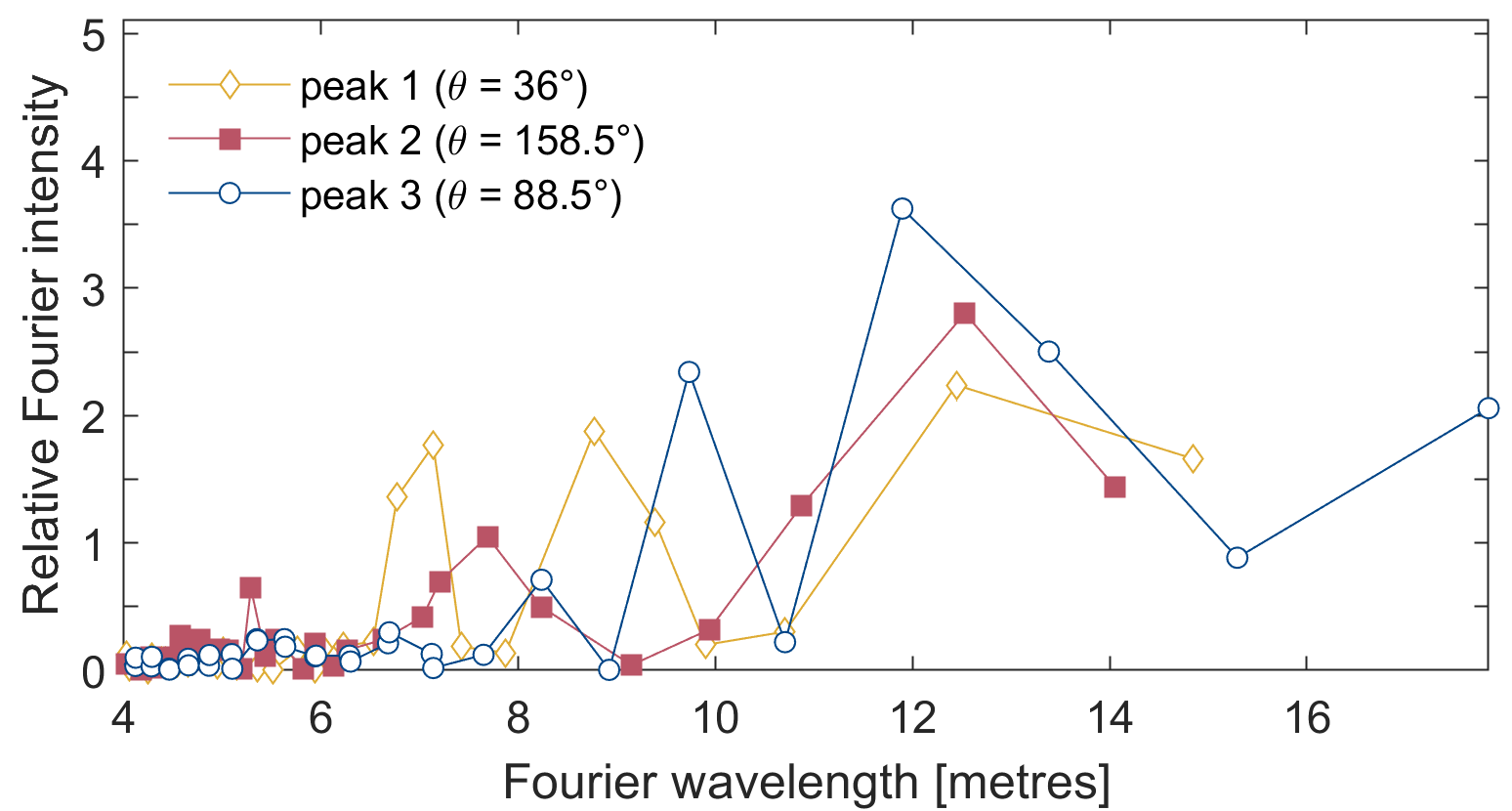}
	\caption{Fast Fourier Transform power spectra along the directions of the three intensity peaks in \autoref{fig:intensities_bar}, obtained by plotting the Fourier modes' wavelength against their intensity. Wavelengths where the modes have great intensity contribute most strongly to the frequency-makeup of the image, i.e., they represent a frequently occurring spacing of linear structures. To represent this relationship more clearly, the wavelengths are converted to metres, using the mean spatial resolution of the image (1.06 m/px). Wavelengths below 4\,m (noise) are cropped from the figure for clarity.}
	\label{fig:powerspectrum}
\end{figure}

\subsubsection{Maxima in the spatial distribution of FFT intensities} \label{sec:wavelengths}

In order to determine the dominant wavelength in the image, the power spectrum of the frequency domain is plotted along the peak direction, where the wavelength $\lambda(i,j)$ is inversely proportional to the mode's radius $r(i,j)$ such that
\begin{equation}
\lambda(i,j) = w \cdot \frac{1}{r(i,j)}
\end{equation} \label{eq:lambda}
where $w$ is the width of the image in pixels. The power spectra for the three highest peaks are shown in \autoref{fig:powerspectrum}. A maximum in the power spectrum corresponds to the Fourier component with the strongest contribution to the structure of the image in the respective direction. If that structure were strictly harmonic, a single maximum would describe exactly its wavelength. If the range of length scales contributing to the structure is not too large, the maximum can still be expected to describe its dominant wavelength.

\subsubsection{Adjustable use-case parameters}

The main human and thus somewhat subjective input in this method lies in determining a combination of method parameters that produces a significant and reliable detection of the relevant structures in the image. Important parameters are the size of the original image, the width of the angular sectors, the values for $r_{min}$ and $r_{max}$, as well as the parameters for the peak-fitting-algorithm used on the angular distribution of FFT intensities. 

These parameters need to be found empirically for each use case, as they depend both on the target feature (e.g. layering, jointing) and the landscape and lighting conditions of the used image. The selection criteria are described in more detail in \autoref{sec:testcase}.

\subsection{Application of method to larger images} \label{sec:largerimages} 

The strength of our method lies in applying it to larger images. This is achieved by dividing the larger image into square sub-images, on which the algorithm operates as described in \autoref{sec:fouriertransform}. The result is a map of directions of geomorphological structures (e.g. lineaments), where each sub-image is replaced with the central area of its backtransformation along the detected direction of the structures (\autoref{fig:results_lay}B). Only the central $\sim$ 50\% of the area of each sub-image provides sensitive results during FFT due to windowing (\autoref{fig:windowing}). The loss in sensitivity is compensated by overlapping the sub-images by 50\% in horizontal and vertical directions, such that the sensitive parts of the sub-images cover the image area in a dense, coherent pattern. 

To analyse a larger image, an appropriate frame size needs to be chosen. The frame size is defined as the length of the edges of an individual, square sub-image in pixels and depends on the features which are to be analysed. 

On the one hand, the frame size should be chosen as small as possible, in order to maximise the resolution of the resulting map and to reduce the uncertainty of the results (e.g. the directions of lineaments) for each individual sub-image. Small but numerous sub-images also improve the detection of trends within the whole image. 

On the other hand, sub-images must be sufficiently large to fit a substantial portion of the feature, e.g. a segment of a layer edge, especially since only the central 50\% of a sub-image's area is effectively processed by the FFT due to windowing. It also should be considered that the frame size limits the range of wavelengths that can be detected in the image ($\lambda \leq 0.5\: \times$ the frame size). 

A frame size with an odd number of pixels (e.g. $101 \times 101$ pixels) is computationally beneficial, as the sub-image's central pixel is unambiguous in this case.

\subsection{Method test case: Hathor cliff on comet 67P}\label{sec:testcase}

\begin{figure} 
    \centering
    \includegraphics[width=\linewidth]{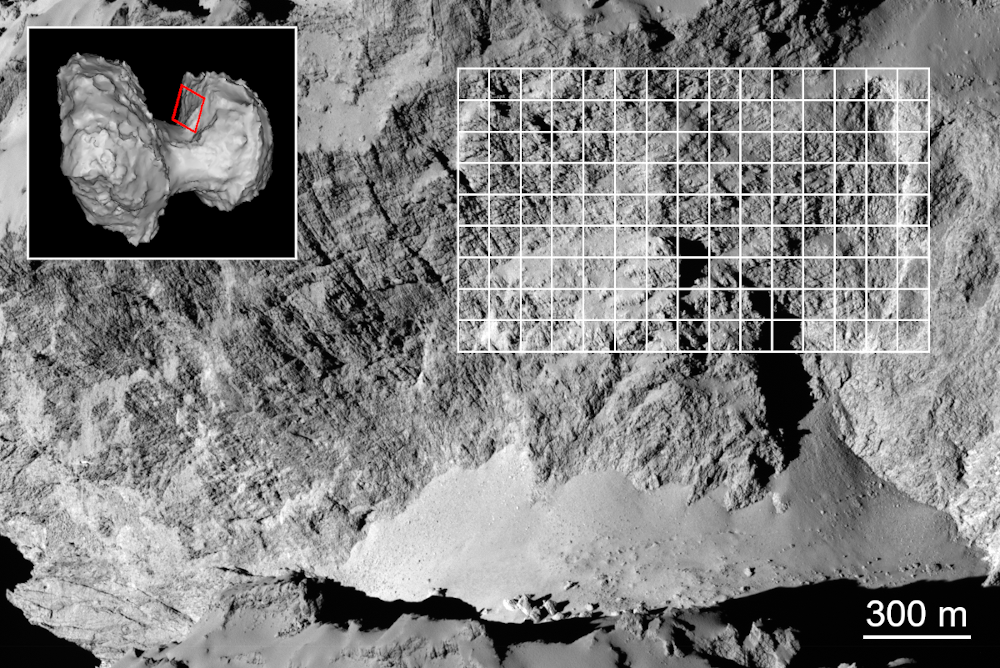}
    \caption{The Hathor cliff on comet 67P, image cropped from OSIRIS image N20140828T124254563ID30F22. On the inset, the red quadrangle indicates the location of Hathor on a shape model of the comet nucleus. The white grid lines indicate the study area used for the method test case (cf. \autoref{fig:results_lay}A).}
    \label{fig:crop_location}
\end{figure}

The Hathor cliff on comet 67P is located in the neck region of the nucleus, on the small sub-nucleus side facing its larger counterpart (\autoref{fig:crop_location}). It is described as a region with complex surface features, including various lineament signatures of layerings and furrows \citep{thomas_2015_morphological,basilevsky_geologic_2017,belton_origin_2018}. The cited publications identified lineaments on the basis of visual inspection. Our initial attempts at quantitative retrieval of information on the layering and other lineaments of the cliff through existing methods was unsuccessful, due to the lack of mappable terraces (used by \citet{massironi_two_2015,penasa_three_2017}) or curved lineaments (used by \citet{ruzicka_mnras_2018}). However, \citet{belton_origin_2018} have made a human-guided attempt to explore layering properties along a 1-dimensional trace through the Hathor cliff. 

For our analysis we selected an image of the Hathor cliff taken by the Rosetta OSIRIS camera, that has solar illumination and camera viewing conditions favourable for identifying geomorphological signatures in the landscape. The image is available from the ESA Planetary Science Archive \citep{besse_psa_2018} with image ID\,=\,N20140828T124254563ID30F22 and includes full geometric and radiometric corrections (calibration level CODMAC 3). The spatial resolution in the image varies between 1.0549 and 1.0619 m/pixel. The viewing angles of the surface elements in the shape model of the cliff ('emission angles') show a maximum at 45$\degr$ (mean value of 37.7$\degr$), with a vast majority of values below 50$\degr$ \citep[Figure 3.21]{ruzicka_thesis}.

From this image, we chose a rectangular area of approximately 800 $\times$ 500 metres, which we split up into square sub-images of size 101 $\times$ 101 pixels (white grid in \autoref{fig:crop_location}).

\begin{figure} 
	\centering
	\includegraphics[width=\linewidth]{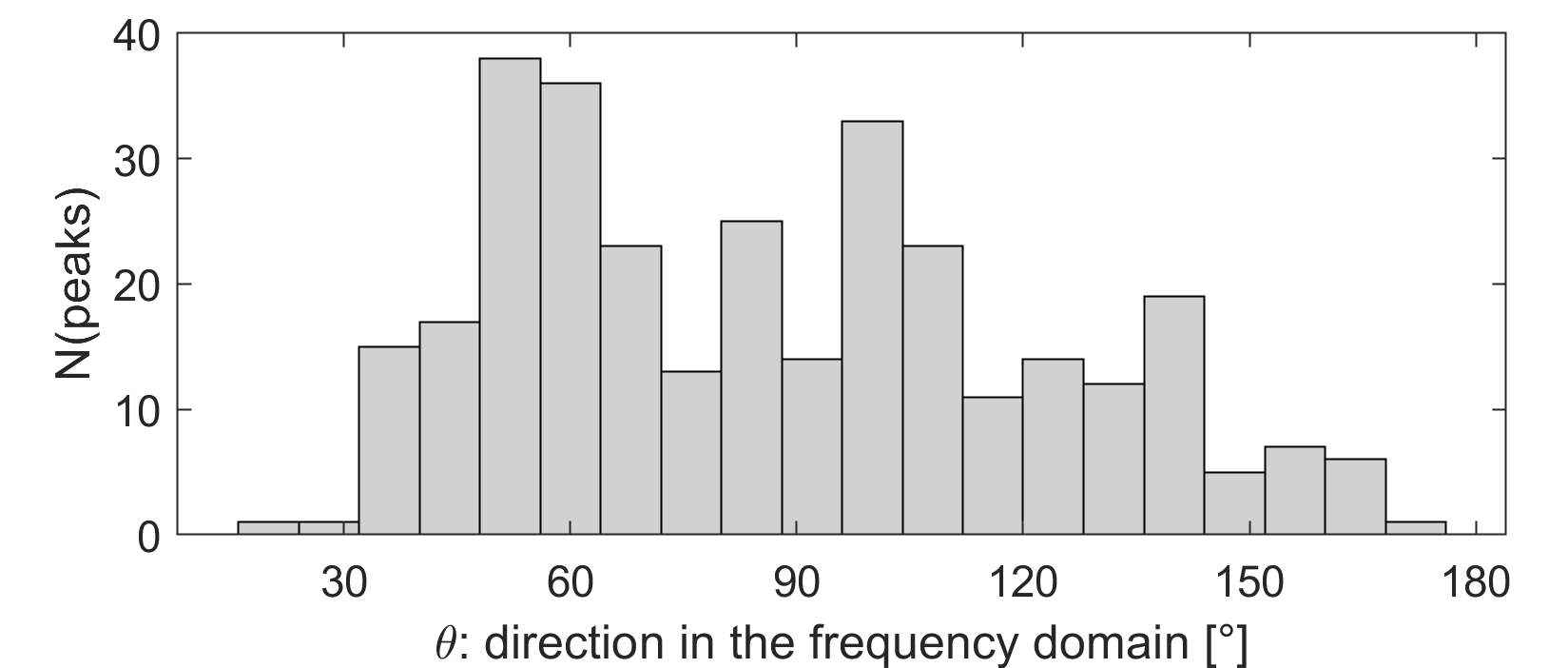}
	\caption{Distribution of intensity peaks by $\theta$ angle for \autoref{fig:results_lay}A. The histogram considers the 3 highest peaks in the intensity spectra of each of the 135 sub-images.}
	\label{fig:histogram_theta}
\end{figure}

Images of geological sites such as this generally contain lineaments in more than one orientation, e.g. layering, jointing, non-bedrock material boulders form local landslides, as well as shadows whose direction depends on the morphology and the direction of sunlight. To constrain orientations of interest, we therefore first took inventory of the directions of all lineaments present in the image. In the study area, lineaments are most frequently found in the directions $\theta \approx 50\degr$, $\theta \approx 110\degr$, and $\theta \approx 140\degr{}$ (\autoref{fig:histogram_theta}). Considering the morphological situation on the Hathor cliff (for instance described in \citet{basilevsky_geologic_2017}), a high probability exists that these directions correspond to the downslope lineaments (furrows, 50\degr), the direction of shadows (110\degr), and the layering-associated lineaments (140\degr). Based on this, we identified a range of acceptable orientations for the sub-horizontal layerings (140\degr{} $\leq \theta \leq$ 165\degr) and for furrows (40\degr{} $\leq \theta \leq$ 70\degr{}).

As mentioned in \autoref{sec:fouriertransform}, a suitable combination of method parameters needs to be selected for each target image. For the Hathor cliff, we experimentally determined that the parameters listed in \autoref{tab:parameters} deliver reliable results for furrow- and layering-associated lineaments. We will now briefly discuss the criteria for their selection and aspects to be considered for choosing sensitive parameter values.

\begin{table}
	\centering
	\caption{Image-specific parameters for the Hathor cliff, determined empirically.}
	\label{tab:parameters}
	\begin{tabular}{ll}
		\hline
		Parameter & Value \\
		\hline
		Frame size              & 101 $\times$ 101 pixels\\
		Sector width            & 5\degr{}\\
		Minimum peak height     & 1.5 $\times$ average intensity\\
		Minimum peak width      & 5 sectors\\
		Minimum peak distance   & 15\degr{}\\
		Minimum radius $r_{min}$ & 6 pixels\\
		Maximum radius $r_{max}$ & 40 pixels\\
		\hline	
		\end{tabular}
\end{table} 

\textbf{Sector width.} Choosing smaller sectors increases the angular resolution of the intensity spectrum (\autoref{fig:intensities_bar}), reduces the uncertainty in the results for the angular orientation of the lineaments, and yields sharper backtransformed images as a smaller angular range of frequencies is considered for the backtransformation. However, as naturally occurring lineaments are rarely exactly planar, the sectors need to be at least wide enough to allow for some variation in direction of lineaments within a sub-image. The sector width has an impact on the choice of the minimum radius $r_{min}$, and it limits the maximum wavelength of repetitive surface features (see below).

\textbf{Minimum peak height.} Peak identification is strongly influenced by the minimum required intensity, or 'height', of a sector. The intensity spectrum of a sub-image showing noisy surface texture instead of aligned linear features fluctuates only slightly around the mean. Squaring the FFT intensities enhances the peaks relative to the noise, but runs the risk of creating false positives from slightly-above-mean intensities. Setting a minimum peak height excludes these false positives. Since illumination varies between the sub-images, it is defined not as a fixed intensity value, but rather as a factor by which a peak needs to surpass the average intensity of all sectors. A suitable factor needs to be determined by trial-and-error for each use case. It will depend largely on the lighting conditions and the surface texture. 

\textbf{Minimum peak width.} Setting a minimum peak width reduces the risk of erroneous peak identification in noisy spectra by excluding outlier values. If the value is chosen too large, valid peaks might be dismissed if their flanks are not steep enough and thus broader than the cutoff value. For the Hathor wall, a minimum width of 5 sectors was found to be most helpful, which, accounting for sector overlap, corresponds to 12.5\degr{}.

\textbf{Minimum peak distance.} Linear features on the Hathor wall are found to run in one of three directions, which are separated from each other by approximately 30\degr{} or more. In the intensity spectrum of some sub-images, potential peaks are located much closer to each other than this. Not both of those potential peaks can be representative of a true feature direction and one peak must be chosen. By selecting a minimum peak distance, the higher one of two adjacent peaks will be selected by the peak-fitting algorithm.

\textbf{Minimum and maximum radius.} The range of wavelengths that can be detected in a sub-image depends on which modes in the FFT are considered. As the radius of a mode is inversely proportional to the wavelength it represents, $r_{max}$ limits the minimum detectable wavelength. This reduces noise by modes which do not contribute strongly to the sub-image anyway. The minimum radius $r_{min}$ determines the maximum wavelength and needs to be set because close to the image centre pixel size exceeds sector width, which would lead to misattributions.

\section{Results and Discussion of method}\label{sec:results}

\begin{figure*} 
    \centering
    \includegraphics[width=0.96\linewidth]{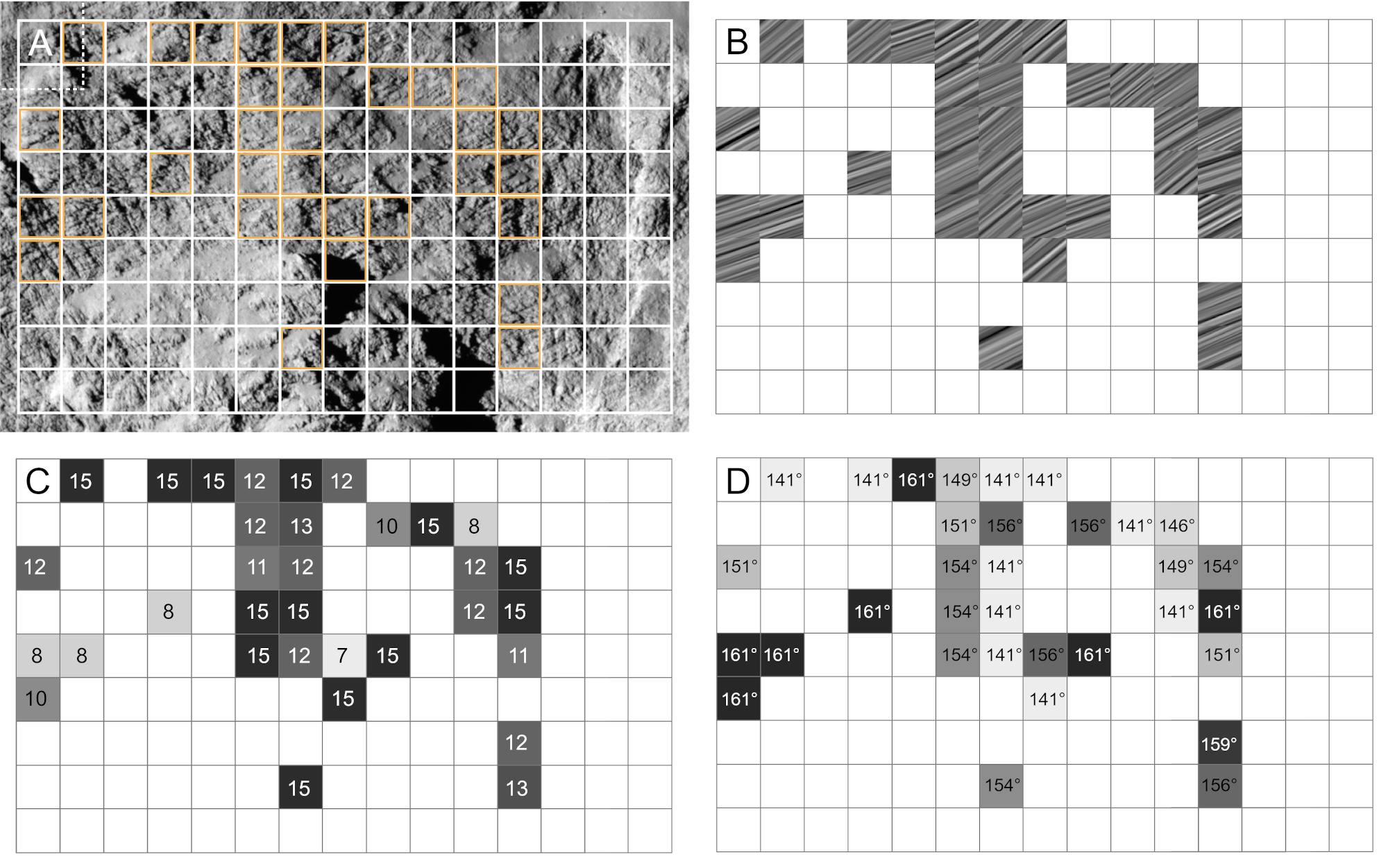}
    \caption{Results for the detection of layerings in the test case of the Hathor cliff. The algorithm used the parameters in \autoref{tab:parameters} and was sensitive to features oriented $140\degr{} \leq \theta \leq 165\degr{}$. \textbf{A:} Image with superimposed grid of sub-images, where each grid element represents the central $\sim$ 50\% of the area of a sub-image (cf. \autoref{sec:largerimages}). The full area of the first sub-image is indicated by the dashed line around the first grid element. Sub-images where layerings were detected are highlighted in orange to guide the eye. \textbf{B:} Map of positive detections of layering, where each sub-image is replaced with the central area of its backtransformation along the detected direction of the layerings. \textbf{C:} Most probable spatial wavelength of layering-associated lineaments in each sub-image (proxy for 'layer thickness'), given in metres, rounded to the nearest integer. Darker fields indicate larger wavelengths. \textbf{D:} Orientation $\theta$ of the layering-associated lineaments in each sub-image, rounded to the nearest integer. Darker fields indicate values closer to the horizontal.}
    \label{fig:results_lay}
\end{figure*}

\begin{figure*} 
    \centering
    \includegraphics[width=0.96\linewidth]{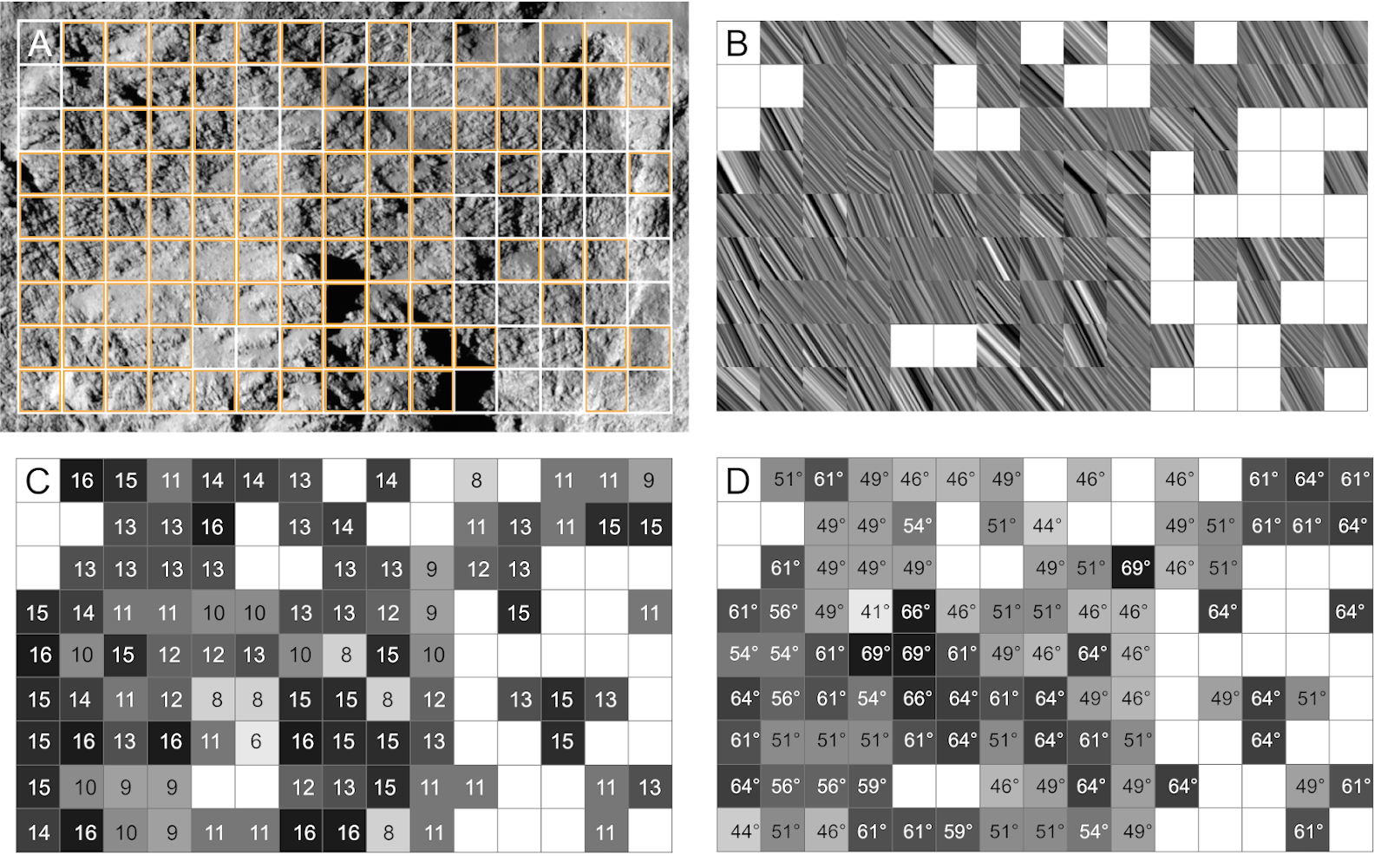}
    \caption{Results for the detection of furrows in the test case of the Hathor cliff. The algorithm used the parameters in \autoref{tab:parameters} and was sensitive to features oriented $40\degr{} \leq \theta \leq 70\degr{}$. \textbf{A:} image with superimposed grid of sub-images (cf. \autoref{sec:largerimages}). Sub-images where furrows were detected are highlighted in orange to guide the eye. \textbf{B:} Map of positive detections of furrows, where each sub-image is replaced with the central area of its backtransformation along the detected direction of the furrows. \textbf{C:} Most probable spatial wavelength of furrow-associated lineaments in each sub-image (proxy for 'furrow spacing'), given in metres, rounded to the nearest integer. Darker fields indicate larger wavelengths. \textbf{D:} Orientation $\theta$ of the layering-associated lineaments in each sub-image, rounded to the nearest integer. Darker fields indicate values closer to the vertical.}
    \label{fig:results_furrows}
\end{figure*}

Using the FFT method described here, we have identified quasi-periodic signatures on the Hathor cliff. The results shown in \autoref{fig:results_lay} provide the information on the orientation (position angle in the image) and the most probable spatial wavelengths of identified sub-horizontal layering-associated lineaments in the various sub-images of the image raster. Results for the vertical furrow lineaments are shown in \autoref{fig:results_furrows}. The results are obtained in a repeatable and comparable way by an automated procedure. 

\begin{figure} 
	\centering
	\includegraphics[width=\linewidth]{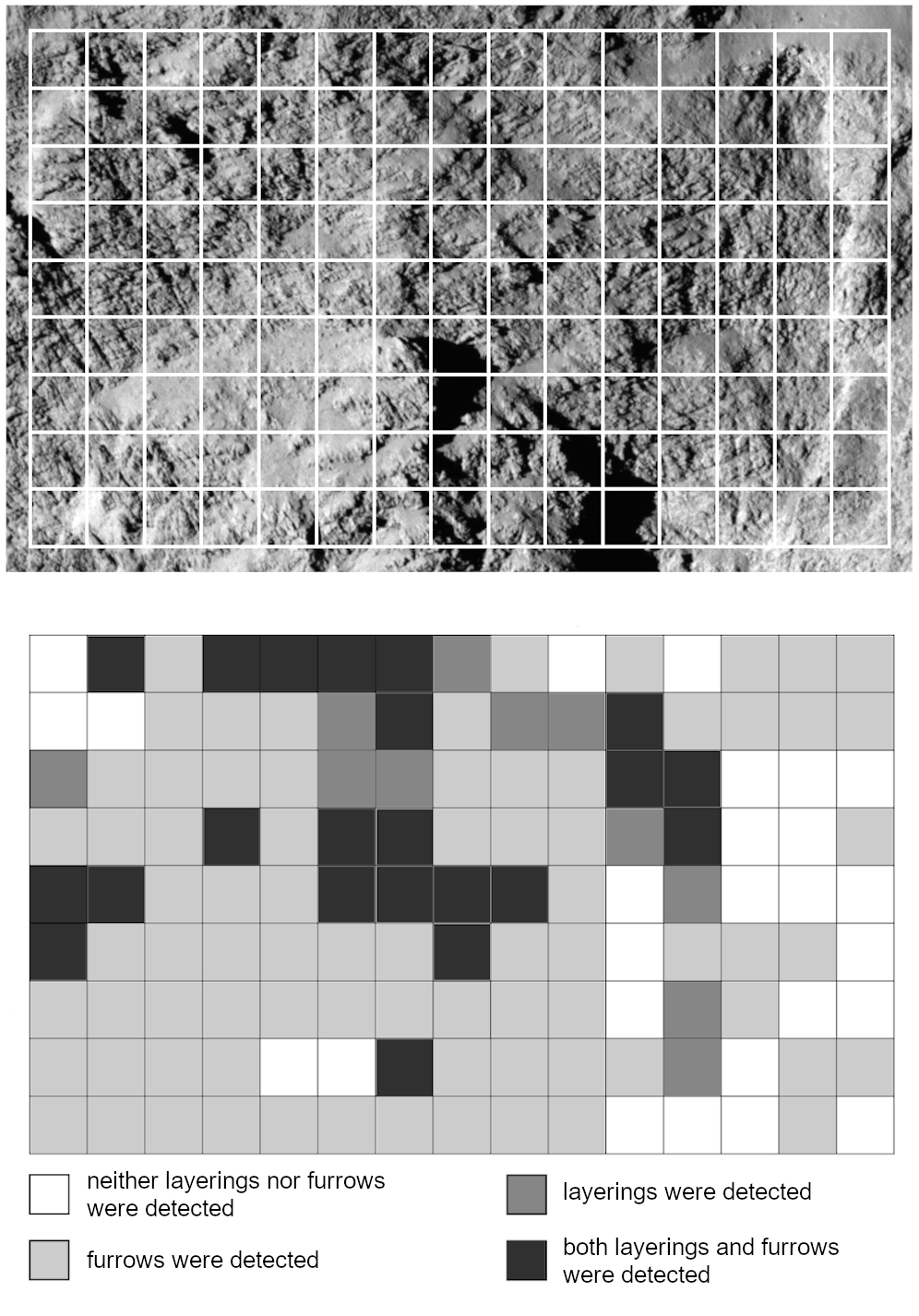}
	\caption{\textbf{Top:} Target area on the Hathor cliff, superimposed white grid indicates the sub-images. \textbf{Bottom:} Map of peak detections on this image, classified by result of lineament detection. This figure is simplified and does not include the overlapping sub-images.}
	\label{fig:peakmap}
\end{figure}

The identified quasi-periodic signatures can be assigned to sub-horizontal and vertical lineaments, i.e., to morphological layerings and furrows, respectively. In addition, various sub-images show substantial areas of high contrast due to local morphology. The identification of the morphological features is done via comparison of the backtransformed FFT image with the respective sub-image of the original image of the Hathor cliff. Various sub-images provide simultaneously information on combinations of the signal patterns, i.e., on layerings together with furrows, on layerings or furrows together with solar illumination signatures, or on all three signal patterns in the same sub-image. This is depicted in \autoref{fig:peakmap}. 

We note that the FFT method does not identify geomorphological signatures (layerings, furrows) in some sub-images where human visual inspection argues for their presence. Increasing the detection sensitivity, for instance by reducing the limits for maximum detection in the angular pattern of the FFT, puts the maximum detection quickly to the noise floor, making the signature identification questionable. On the other side, the FFT method is capable to synthesise quasi-periodic signatures in sub-images with complex brightness distributions, where the human visual inspection experiences difficulties in pattern recognition. 

We have estimated a typical uncertainty of about 3\degr{} for the results on the orientation of quasi-periodic structures in the sub-images; for the wavelength estimations an uncertainty of 1-2 pixels should be assumed (for the pixel range 5-10 and above). In the following we describe separately results for layering and furrow signatures.

\subsection{Layerings}\label{sec:layerings}

Layering structures were identified in 33 of the 135 panels of the analysed Hathor cliff region (\autoref{fig:results_lay}). Positive detections are spread across the analysed region without evident clustering. It is noted that no layerings were found by the FFT analysis in the area of the landslide and at the right edge of the image where changes in the geomorphology of the landscape are seen (\autoref{fig:results_lay}A). The possible interconnection of layerings in adjacent sub-images is not addressed by this analysis.

The orientation of the layering edges is $141\degr{} \leq{} \theta \leq{} 161\degr{}$ in the Fourier domain (mean $\theta = 151\degr{}$), corresponding to $51\degr{} \leq{} \alpha \leq{} 71\degr{}$ in the image of the Hathor cliff (\autoref{fig:results_lay}D). Thus, the layering edges run roughly perpendicular to the current orientation of the local gravity vector at the Hathor cliff (personal communication with S. Hviid, DLR Berlin-Adlershof). As the formation of the layerings predates the merging event of the two nucleus lobes \citep{massironi_two_2015,penasa_three_2017}, their orientation within the current gravity field may not provide new evidence for their formation process.

The maxima of the wavelength distribution (i.e., spacing) for the layerings are between 7.0 and 15.9 pixels (mean $\lambda$ = 12.9 pixels). The corresponding distances on the surface of the comet (in metres) are approximately equal to these values, as the ground resolution for this image averages 1.06\,m/px.

\cite{belton_origin_2018} give a typical spacing for layerings in the Hathor cliff of 14\,m, close to the result we found. Their value was obtained by Fourier analysis of a brightness profile through the Hathor cliff, located within the right half of our study area. Their profile is one-dimensional, roughly vertical through the cliff, but it is not following a straight line and it is selected by eye covering a section of the cliff where layerings are well discernible. The advantage of their 1-dimensional profile is that it covers a long, continuous cut through exposed layerings in the Hathor cliff and thus it delivers a strong Fourier signature for the quasi-periodic layerings present therein. The FFT approach chosen in this paper works on discontinuous structures and keeps information on their context (location and orientation). Moreover, it is an automatic procedure and is applicable to different surface signatures in a single step and without human interaction.

In our analysis the extension of sub-images with identified layerings is $\sim$ 600 m horizontally and $\sim$ 400\,m vertically. Other studies that use eye-guided identifications of the layerings found comparably large values for the spatial spreading of layerings, e.g. Figure 2 of \citet{basilevsky_geologic_2017} and Figure 2 of \citet{belton_origin_2018}.

\subsection{Furrows}\label{sec:furrows}

In the analysed region of the Hathor cliff, detections of furrows are more abundant than of layerings (in 89 out of 135 sub-images; see \autoref{fig:results_furrows}). This is because, on average, they are more pronounced in their FFT signatures and produce stronger peak factors (i.e., the ratio between maximum and mean level in the angular distribution of the sum of FFT intensities) compared to the layerings. 

The spatial density of furrows is higher in the left hand side of the analysed image (see \autoref{fig:results_furrows}B). This includes the region of the landslide, tentatively identified by \citet{basilevsky_geologic_2017}.

The distribution of orientations in the Fourier domain show maxima for $41\degr{} \leq \theta \leq 69\degr{}$ (mean $\theta$ = 54\degr{}), which corresponds to an $\alpha$ range of -49\degr{} to -21\degr{} in the original image. The typical spatial wavelengths of the furrows, between 6.1 and 15.7 pixels or metres (mean $\lambda$ = 12.4), is notably similar to the spacing of layerings. This is contrary to the visual impression of the image, but caused by the fact that two semi-parallel sets of furrows are present in the image. One set is broader and darker and therefore dominant in the visual impression, although they are less abundant in the image. The other set is narrower and lighter and therefore draws less visual attention, even though they are more abundant. The darker furrows are spaced further apart than the lighter ones. This results in the overall visual impression that the furrows are spaced further apart than they generally are.

In this study, the chosen parameters (\autoref{tab:parameters}) were optimised for the detection of layerings, not furrows. Particularly the minimum radius $r_{min}$ limits the maximum detectable wavelength. At $r_{min} = 6$ px, the maximum detectable wavelength is $\lambda = 16.6$ px (\autoref{eq:lambda}). The detection of wavelengths for the furrows could be improved by choosing a smaller minimum radius, as well as a larger frame size to accommodate features with greater spacing. Either option would, however, negatively impact the results in other ways (cf. end of \autoref{sec:testcase}).

The local gravity vector is oriented downhill approximately parallel to the cliff wall. This makes the interpretation of the furrows as downslope signatures of boulders and cliff material very likely \citep{basilevsky_geologic_2017}. Since no layering lineaments are identified in the region of the landslide (see \autoref{sec:layerings} below), it appears plausible that the furrows in this surface area were formed after the landslide took place. Based on this superposition, the layerings are a geologically older structure than the furrows.

\section{Discussion of the formation of layerings and furrows on comet 67P}\label{sec:conclusions}

We observe that layering-associated lineaments are exposed at considerable spatial extent on the nucleus surface of comet 67P, as they are detectable on the whole Hathor cliff. The cliff wall represents a slice plane that exposes the nucleus interior to a significant radial depth \citep[cf.][Figure 6A]{penasa_three_2017}. From this we conclude that layerings are not only a surface feature on the nucleus of comet 67P, but rather permeate at least several hundred metres below the surface of the small lobe. In combination with observations discussed in \citet{massironi_two_2015}, \citet{penasa_three_2017}, and \citet{ruzicka_thesis}, the same appears to be true for the large lobe. We therefore conclude that the entire nucleus of comet 67P has a pervasive layered internal structure.

We further observe that on the visible exposure of the Hathor cliff, there is little variation in orientation and spacing of the layerings, particularly with regard to radial depth. 

We now discuss the compatibility of our observations with the two main models of formation for layerings in cometary nuclei.

\subsection*{Layers as primary structures}

In the first model, the layers we observe today were formed in the nucleus structure during accretion as a so-called primary structure. On Earth and other terrestrial bodies, these type of layers occur in most sedimentary and igneous rocks, glaciers, and any other materials that are deposited sequentially. Strata in a sequence of primary layerings may be distinguishable from each other by variations in grain size, colour, or composition. Layering boundaries in primary structures can also result from pauses in deposition that allow the older deposits to undergo changes before additional material covers them. The nature of primary layerings could thus be considered 'sedimentary', and their creation would require periodic variations of either material or conditions during accretion. 

In order to align our observations of the layerings on comet 67P with the model of primary layerings formed during accretion, several assumptions need to be made.

First, in order to form layerings such as those on the Hathor cliff, that extend laterally at constant thickness and have smooth, parallel layer boundaries, a steady depositional environment would be required. The environment would need to have had a uniform direction of material transport (as we do not observe cross-bedded layers) and few to no collisions (as we do not observe significant local denting or disruptions of the layers). Additionally, an explanation must be found for the lack of variation in layering thickness with radial depth, as exposed on Hathor. Of particular interest is the absence of a trend in layer thickness, which would require an increasing mass-accretion rate over time, as a constant mass-accretion rate on an approximately spherical or elliptical body would result in decreasing layer thickness with increasing radius of the body. If, however, cometary nuclei are not formed as hierarchical rubble piles \citep[e.g.][]{davidsson_primordial_2016} or streaming instabilities \citep{blum_pebblecollapse_2017}, but instead by a - perhaps even recent - collapse of a pebble cloud after a collision event \citep{jutzi_subcatastrophic_2017}, a mechanism must be found that created layer boundaries in a rapidly collapsing particle cloud.

\subsection*{Layers as secondary structures}

In the second model, the layers were superimposed onto an originally homogeneous nucleus some time after accretion. These layerings are then called secondary structures. An example from Earth is the stratification in soils, where layers are developed during pedogenesis by biochemical and physical vertical transport processes within the material. A mechanism for secondary layer formation in cometary nuclei was recently proposed by \citet{belton_origin_2018}, in which a primordial and mostly homogeneous rubble-pile nucleus is transformed into a layered body by the self-sustaining, dual-mode propagation of amorphous-to-crystalline water ice phase-change fronts from the surface into the nucleus interior. They hypothesise that the direction of propagation of the fronts is controlled by the radial outflow of carbon monoxide and a coarse layered structure in the primeval material below the front, the origin of which remains to be explained. They find that for a nucleus of the size of 67P, the fronts would have completed the conversion within approximately 2000 years. It is noted that the trigger for the phase change may be of regional extension and still has to create layers of global and quasi-concentric nature in the nucleus.

\citet{belton_origin_2018} expect this mechanism to create geologically young layerings of global extent with an approximate radial thickness of 14\,m, which converges with our findings. 

In contrast to the primary-structure model, this model places fewer requirements on the external conditions in the Solar System. Once a comet reaches the 'Centaur phase' in its life cycle (i.e., when gravitational interactions with the outer planets have directed the comet into an orbit much closer to the Sun), the nucleus surface will eventually reach a critical temperature that initiates the phase-change reaction. From this point on, the process progresses without the need for further external input. The periodicity represented by the layerings may be created through alternating fast and a slow phases of the thermal reaction in the phase-change front. It can also be supported by a periodicity in the heat wave intruding into the nucleus during the passage of the perihelion arc of the orbit.

\subsection*{Furrows}

The second set of linear structures on the Hathor cliff have a pronounced negative topography and are oriented roughly parallel to the current local gravity vector on the cliff face. We called these structures 'furrows' because their orientation suggested that they were carved into the weak nucleus material during the landslide event that exposed the cliff face as we see it today. However, the remnant of the landslide mass that is resting at the foot of Hathor (\autoref{fig:crop_location}) is too small to have carved the lineaments in the whole cliff. This contradiction could be explained by the consideration that the body of the landslide mass, having been internally stressed and fragmented while it slid several hundred metres down the cliff face and finally collided with the neck region at the bottom of the cliff, would have been eroded at a much faster pace than the surrounding, structurally more intact material. The talus of fine-grained material that surrounds the landslide mass (\autoref{fig:crop_location}) may very well be the erosional remains of what is missing of the landslide mass. It is harder to explain the observation that 'furrows' are visible in most of the remaining landslide mass. Either those lineaments were left behind in the material by an earlier, additional cliff collapse (which eludes our study), or an alternative hypothesis must be found for the origin of the 'furrows'.

One possibility is that, as the cliff continues to erode, pebbles and boulders detach from the top and leave marks in the material as they tumble down the cliff. An argument in favour of this hypothesis is the orientation of the lineaments along the gravity vector. Another argument is that this mechanism would also leave marks indiscriminately on both cliff face and landslide mass. On the other hand, it seems unlikely that this might result in furrows of such apparent depth. Moreover, one must wonder where these boulders ended up, as only a small number of them can be seen at the foot of the cliff. By contrast, the foot of the freshly collapsed Aswan cliff is strewn with debris and countless boulders \citep{pajola_pristine_2017}. Perhaps, given the conditions in the neck region (including gas outflow and dust transport, e.g. \citet{lai_gasoutflow_2016}, or shearing forces, e.g. \citet{matonti_bilobate_2019}), sufficient time has passed since the landslide event at Hathor for boulders to have been covered, eroded, or cleared away from the foot of the cliff.

A second option is that repeated material ejection from the adjacent Hapi region could be locally eroding the cliff and carve the furrows. \citet{shi_hapi_2018} have shown that some of the gas-and-dust-flows called 'jets' that originate from Hapi can intersect with Hathor, at angles and height-to-width ratios compatible with the furrows we observed. 
While we do not rule out that this mechanism contributes to the furrows on Hathor, we are doubtful that it is their main source. 
For one, the gas jets were shown dissipate with increasing distance from their source, such that their contours dissipate at about 800\,m above the ground \citep{shi_hapi_2018}. The densely furrowed top of Hathor, however, is approximately twice as far removed from the closest point of Hapi. We therefore argue that the jets could not have reached this area with sufficient force or focus to carve into the material.
In addition, the jets are described as occasional, short-lived outbursts on unpredictable trajectories. Thus they can hardly be expected to provide the sustained, localised stream of material required to carve furrows that appear to be several metres deep. 
Finally, a process would be required to guide the jets such that they do not carve into Hathor in a random orientation, considering the pronounced parallel orientation of the furrows as well as their alignment with the current gravity vector.  

A third possibility is that the 'furrows' are not mechanical tracks at all, but rather thermal expansion fractures created as the cometary material expands when it approaches perihelion. However, while temperature-related fractures were observed on and modelled for comet 67P by several authors, they were mainly reported to present themselves as irregular fracture networks \citep[e.g.][]{elmaarry_fractures_2015,attree_thermal_2018} or in a polygonal configuration \citep{auger_polygons_2018}, not in a sub-parallel pattern.

\section*{Acknowledgements}

BKR gratefully acknowledges support and funding by the International Max-Planck Research School (IMPRS) for Solar System Science at the University of G\"ottingen, where this research was conducted as part of a PhD thesis \citep{ruzicka_thesis}. We thank Luca Penasa for the fruitful discussion on models of layering formation. Stubbe Hviid, DLR Berlin-Adlershof, provided very useful information on the local gravity field at the Hathor cliff on comet 67P. We thank an unknown reviewer for raising helpful questions about the interpretation of our results.

\section*{Data Availability Statement}

The data underlying this article are available in the article and in its online supplementary material.

For easier reproducibility of our method, we provide a demo code for MATLAB in the supplementary data (available online).



\bibliographystyle{mnras}
\bibliography{bibliography}

\appendix
\section{Application to other geological cases}\label{sec:applications}

The purpose of this Appendix is to demonstrate that our methodology can be applied to various geoscientific targets. It is a powerful tool for detecting and analysing layerings and other linear structures that works on different spatial scales. However, it requires opportune viewing and lighting conditions.

\subsection{Stratification in the 'The Wave' formation, Arizona}\label{sec:wave}

\begin{figure} 
	\centering
	\includegraphics[width=\linewidth]{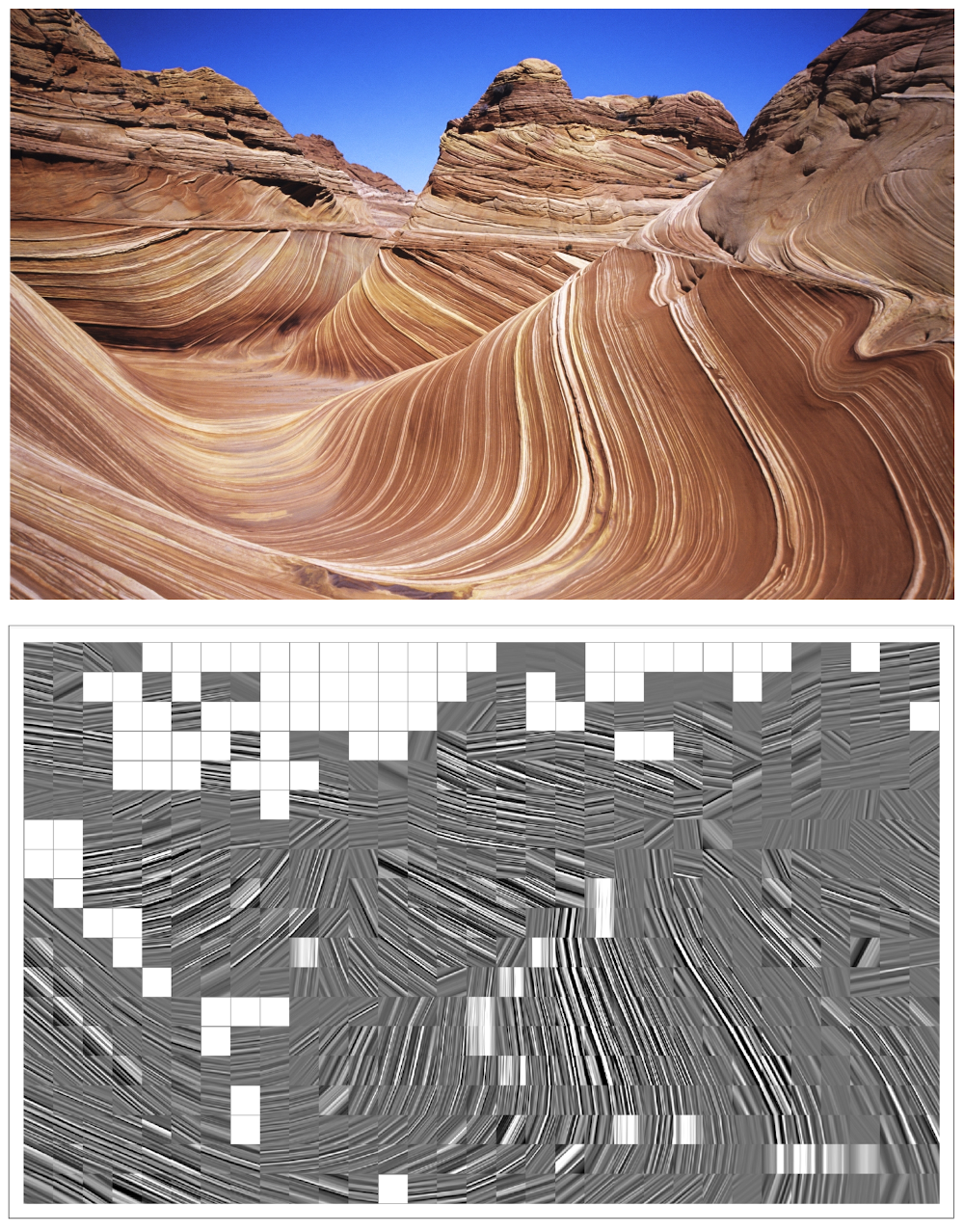}
	\caption{\textbf{Top:} Photograph of the famous sandstone formation called 'The Wave', located in Arizona (Earth). Image available under Creative Commons CC0 1.0 Universal Public Domain Dedication \citep{wiki_thewave}. \textbf{Bottom:} Map of backtransformed sub-images along the direction of the highest peak.}
	\label{fig:thewave}
\end{figure}

\autoref{fig:thewave} shows the result for a colour image of a layered, eroded sandstone formation in Arizona, USA, called 'The Wave'. The layerings differ in colour and hardness due to variations in grain size \citep{freeman_1975_navajo}. The image size is 1601 $\times$ 1001 pixels. The same configuration of parameters that were used for the Hathor cliff also proved useful for this image. 

The map of backtransformations along the direction of the highest peak in the Fourier intensity spectrum (\autoref{fig:thewave}, bottom) clearly reproduces the main stratification visible within the sandstone. Almost all empty frames (i.e., no detections) below the skyline are caused by the Fourier energy not clearing the Minimum Peak Height threshold. Lowering the threshold, however, introduces false detections, as the image areas covered by those frames contain more than one layering orientation, but the algorithm applied was tuned to handle only one peak per tile. The number of empty tiles could be reduced by choosing a smaller frame size. Rightly, no lineaments were detected in the sky.

\subsection{Crystal twinning in a mineralogical thin section viewed in polarised light}

\begin{figure} 
	\centering
	\includegraphics[width=\linewidth]{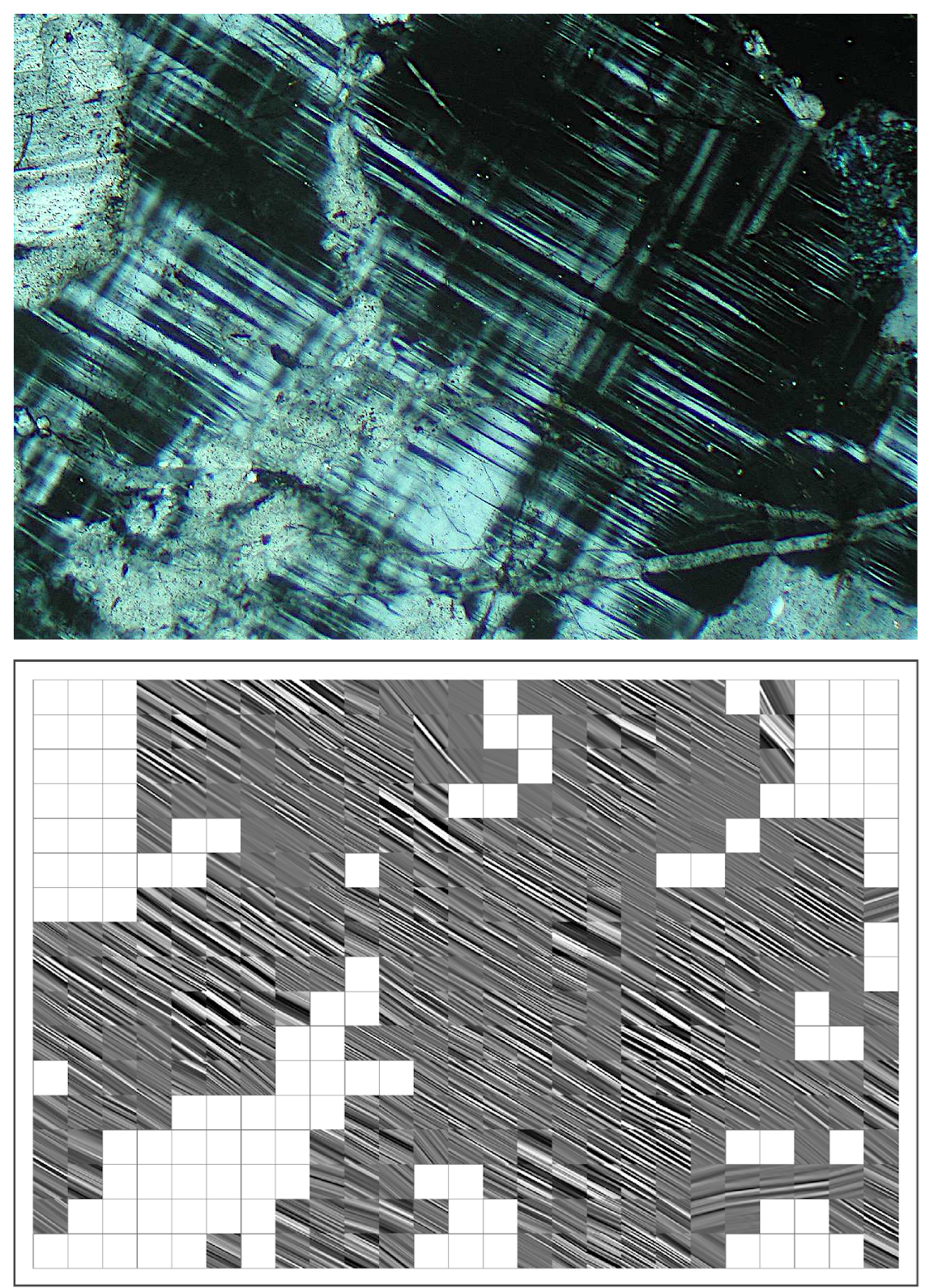}
	\caption{\textbf{Top:} Photomicrograph of a thin section of microcline rock from Earth, viewed in crossed polarized light. The image width is approximately1 cm.  Image  cropped  from  original,  which  is  available  under  Creative Commons License BY-SA 2.5 \citep{ries_2007}. Bottom: Map of backtransformed sub-images along the direction of the highest peak, using a minimum peak height of 3.0×average intensity, frame size 101 px.}
	\label{fig:thinsection}
\end{figure}

The method described in this contribution can also be applied to microscopic structures. We applied the method to a photomicrograph of a thin section of a microcline rock from Earth (\autoref{fig:thinsection}). The image shows feldspar crystals which contain linear internal structures called twinning. Several other linear structures are present at various orientations. Some areas in the image do not contain lineaments. 

The  map  of  backtransformations  accurately  represents  the  orientations of the twinning lines, which means that the sector width was appropriately chosen. The variations in their spacing were found correctly for most of the image. The accuracy of wavelength detection could be improved, at the cost of spatial resolution, by increasing the  frame  size.  In  order  to  focus  the  analysis  on  the  twinning lines  and  reduce  the  effect  of  other  textures,  the  minimum  peak height threshold was set at three times the average intensity in each tile

\bsp	
\label{lastpage}
\end{document}